\documentclass[twocolumn,aps,pra]{revtex4-1}
\usepackage{epsfig,amsmath,graphicx}

\usepackage{bbold}

\newcommand{\ffig}[4]{
\begin{figure}
\centering
\includegraphics[width=#4]{#2}
\caption{#3}
\label{#1}
\end{figure}
}

\newcommand{\ffigDouble}[4]{
\begin{figure*}
\centering
\includegraphics[width=#4]{#2}
\caption{#3}
\label{#1}
\end{figure*}
}

\newcommand{\be}{\begin{equation}}
\newcommand{\ee}{\end{equation}}
\newcommand{\ba}{\begin{eqnarray}}
\newcommand{\ea}{\end{eqnarray}}
\newcommand{\fr}{\frac}

\begin{document}

\title{Full characterization of a highly multimode entangled state embedded in an optical frequency comb using pulse shaping}

\author{R. Medeiros de Ara\'{u}jo, J. Roslund, Y. Cai, G. Ferrini, C. Fabre and N. Treps}
\affiliation{Laboratoire Kastler Brossel, UPMC Univ. Paris 6, ENS, CNRS; 4 place Jussieu, 75252 Paris, France}


\begin{abstract}

We present a detailed analysis of the multimode quantum state embedded in an optical frequency comb generated by a Synchronously Pumped Optical Parametric Oscillator (SPOPO) \cite{Roslund2013}. The full covariance matrix of the state is obtained with homodyne detection where the local oscillator is spectrally controlled with pulse shaping techniques. The resulting matrix reveals genuine multipartite entanglement. Additionally, the beam is comprised of several independent eigenmodes that correspond to specific pulse shapes. The experimental data is confirmed with numerical simulations. Finally, the potential to create continuous-variable cluster states from the quantum comb is analyzed. Multiple cluster states are shown to be simultaneously embedded in the SPOPO state, and these states can be revealed by a suitable basis change applied to the measured covariance matrix. 

\end{abstract}

\maketitle

\section{Introduction}

Photonic architectures have emerged as a viable candidate for the development of quantum information processing protocols. Photons are immune from environmental disturbances, readily manipulated with classical tools, and subject to high efficiency detection \cite{obrien2009photonic}. For these reasons, many proof-of-principle experiments have been demonstrated that utilize either optical q-bits in the discrete variable (DV) regime \cite{walther2005experimental} or fluctuations of the quantized electric field (termed ``q-modes") in the continuous variable (CV) regime \cite{Furusawa2011}. Yet, an interaction among various photonics channels must be established in order to implement a universal set of quantum logical operations (i.e., 2-qubit gates). 

While strong nonlinear interactions at the single-photon level are difficult to achieve, it is possible to initiate an interaction among photonic channels through the act of measurement. 
Such measurement-induced nonlinearities are the basis of linear optical quantum computing \cite{kok2007linear,ralph2010optical}. The KLM scheme of quantum computing \cite{knill2001scheme} utilizes single photon sources, a linear optical network, and introduces the requisite nonlinearity with photon-counting detectors. Although the KLM scheme is fundamentally nondeterministic, it may, in principle, be rendered deterministic with the addition of entangled multiphoton ancilla states. Nonetheless, the overhead necessary to incorporate these states grows rapidly and presents a challenge to the practical scalability of the scheme. 

An alternative approach has recently emerged that exploits the act of projective measurement itself as a means for achieving quantum gates \cite{raussendorf2001one}. In particular, a quantum logical operation can be realized by measuring the state of single nodes contained within a highly entangled multipartite state - the cluster state \cite{Nielsen2006,Lloyd2012}. Due to the multipartite nature of the entanglement, the result of a measurement propagates throughout the cluster in a deterministic fashion. Importantly, different logical gates are implemented by altering only the basis in which individual nodes are measured; consequently, the choice of basis does not necessitate a change in the cluster structure itself. As a result, the primary difficulty for implementing measurement-based computing schemes lies in the generation of the cluster state, which requires large scale entanglement. 

Optical cluster states have been successfully constructed both in the DV \cite{walther2005experimental} and CV \cite{vanLoock2008,Furusawa2011} regimes. Continuous-variable entanglement, which is the domain of the current work, is of particular interest since the electric field is efficiently controlled and measured with classical devices, and the unconditional nature of photon generation allows for both high signal-to-noise ratios and data transfer rates. The traditional methodology to construct CV clusters is to introduce a series of independent
squeezed states of light into a linear optical network that is arranged in such a way as to produce the desired entanglement \cite{vanLoock2007}. Each node contained within these states, however, necessitates its own source of nonclassical states. Consequently, the incorporation of a large number of such modes rapidly encounters a complexity ceiling in terms of scalability and flexibility. 

Alternatively, a multimode source may be exploited in which all of the requisite modes are copropagating within a single beam. One avenue toward cluster state generation exploits temporal encoding \cite{yokoyama2013optical}. Additionally, spatially multimode beams have proven useful for the generation of cluster states when detected with a spatially-resolved, multi-pixel apparatus \cite{Armstrong2012}. However, achieving spatial degeneracy over multiple modes is technically challenging, and an alternative approach is to generate frequency multimode beams, which may be accomplished with optical cavities that are resonant for a large number of copropagating frequency modes. Toward this end, optical frequency combs possess an intrinsic highly multimode structure due to the large number of individual frequencies contained within the comb. The frequency comb has already proven a reliable source of cluster states as the downconversion of a single pump photon in an OPO with a broad phase-matching bandwidth creates sets of entangled q-modes \cite{pfister2011,pfister2008,Chen2013}.  

The present work demonstrates the use of an optical frequency comb to synchronously drive the downconversion process. The result is a highly multimode quantum state of light
that may be described either as a product of uncorrelated non-classical states that span the entire breadth of the downconverted spectrum and have specific pulse shapes, or as a highly entangled multipartite state. We show that the combination of homodyne detection with ultrafast pulse shaping permits recovery of the state's full covariance matrix in a basis of up to eight modes. This description reveals that multiple cluster states are simultaneously present in the multimode beam. 

The paper is organized as follows: Section~\ref{sec:theory} outlines the theoretical principles governing formation of the quantum comb as well as the various bases in which it may be analyzed. The quantum comb is characterized in terms of its covariance matrix. The methodology by which this matrix is obtained is detailed in Section~\ref{sec:methods}, and demonstration of the state's multimode character is presented in Section~\ref{sec:results}. Given the covariance matrix, Section~\ref{sec:clusters} illustrates how the quantum comb may be examined in the various bases that reveal the presence of cluster states. Finally, concluding remarks and an outlook toward future development are discussed in Section~\ref{sec:discussion}.

\section{Theoretical Description} 
\label{sec:theory}

Nonclassical CV photonic states are efficiently generated with an optical parametric oscillator (OPO). In the frequency comb regime, the high peak powers associated with ultrafast pulses elicit a strong nonlinear material response, which, in turn, provides an efficient platform for the creation of highly nonclassical states \cite{wenger2004non}. Moreover, a femtosecond pulse train contains upwards of $\sim 10^{5}$ individual frequency components, and is therefore readily described as a multi-frequency-mode object. The simultaneous downconversion of all these frequency elements in a nonlinear optical element induces an intricate ensemble of both symmetric and asymmetric frequency correlations with respect to the carrier frequency $\omega_0$ that extends across the breadth of the resultant comb~\cite{Pinel2012} (Fig.~\ref{fig-PDC}). These correlations are preserved provided that the optical cavity is synchronously pumped by the laser. If the resonant frequencies of the cavity are written as $\omega_{p} = \omega_{0} + p \cdot \omega_{\textrm{FSR}}$, with $p\in\mathbb{Z}$, this condition implies that $\omega_{\textrm{FSR}}$ is both the cavity free spectral range and the repetition rate of the pump laser. Such a device is called a SPOPO (Synchronously Pumped OPO).

\ffig{fig-PDC}{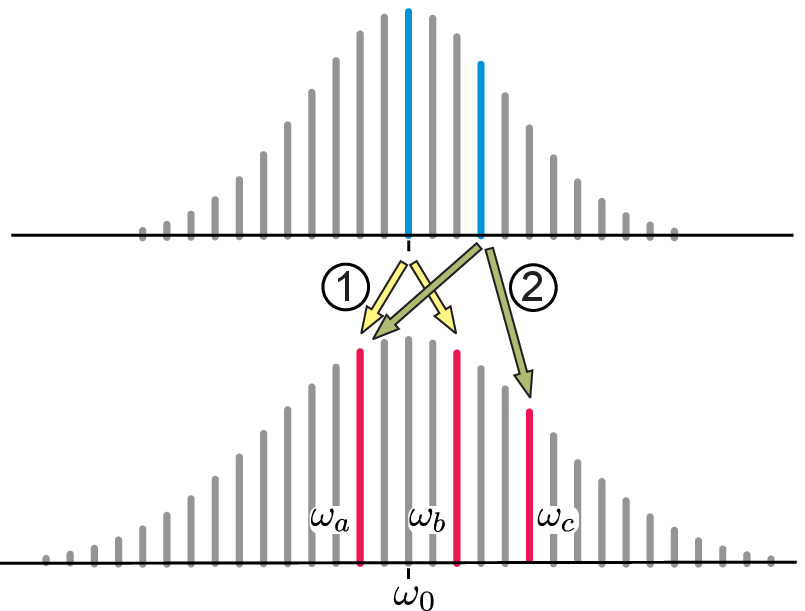}{Parametric downconversion of a femtosecond comb. The splitting of a single pump photon of frequency $2 \omega_{0}$ by pathway $1$ creates entanglement between the frequencies $\omega_{a}$ and $\omega_{b}$. An additional pump photon may downconvert by pathway $2$ and correlate frequencies $\omega_{a}$ and $\omega_{c}$. A correlation is also established between frequencies $\omega_{b}$ and $\omega_{c}$ by virtue of their mutual link to $\omega_{a}$. In this manner, every frequency of the downconverted comb becomes correlated with every other member of the comb.}{65mm}

The Hamiltonian corresponding to a single pass in the crystal that describes the parametric coupling between different cavity modes is then

\begin{equation}
\hat{H} = \textrm{i} \hbar  g \sum\nolimits_{m,n}  L_{m,n}\, \hat{a}_{m}^{\dagger} \hat{a}_{n}^{\dagger} + \textrm{h.c.},
\label{hamiltonian}
\end{equation}
where $g$, proportional to the pump amplitude, regulates the overall interaction strength and $\hat{a}_{m}^{\dagger}$ is the photon creation operator associated with a mode of frequency $\omega_{m}$. The coupling strength between modes at frequencies $\omega_m$ and $\omega_n$ is governed by the matrix $L_{m,n}=f_{m,n} \cdot p_{m+n}$, where $f_{m,n}$ is the phase-matching function \cite{walmsley2001,walmsley2008} and $p_{m+n}$ is the pump spectral amplitude at frequency $\omega_{m}+\omega_{n}$ \cite{patera2010}. In the absence of loss, the evolution of a single mode $\hat{a}_{m}$  is then specified by

\begin{equation}
\frac{d \, \hat{a}_{m} }{dt}   = g \sum\nolimits_{n}  L_{m,n}\, \hat{a}_{n}^{\dagger},
\end{equation}
which reveals that following the downconversion event, each frequency mode is  coupled to every other mode with a strength moderated by $L_{m,n}$. Consequently, the downconversion of an ultrafast frequency comb has the potential to serve as a rich source of multipartite entanglement \cite{Valcarcel12}. 
\subsection{Squeezed Mode Basis}

An alternative description of the state is obtained upon diagonalizing the coupling matrix $L_{m,n} = \sum_{k} \Lambda_{k}X_{k,m}\, X_{k,n}$, where $\{\Lambda_k\}$ and $\{X_k\}$ are its eigenvalues and eigenvectors, respectively \cite{Leuchs2002}. A new set of ``supermodes'' $\hat{S}_{k}$ may be defined that are linear combinations of the original, single frequency modes: $\hat{S}_{k} = \sum_{i}X_{k,i} \, \hat{a}_{i}$. The total Hamiltonian is then written as a sum of single-mode squeezing Hamiltonians independently acting on each supermode \cite{patera2010}:

\begin{equation}
\hat{H} = \textrm{i} \hbar  g \sum\nolimits_{k} \Lambda_{k} \, \hat{S}_{k}^{\dagger \, 2}+  \textrm{h.c.}
\label{sqz-hamiltonian}
\end{equation}
The eigenspectrum $\Lambda_k$ specifies the number of non-vacuum, uncorrelated squeezed states contained in the SPOPO output and their associated degree of squeezing. Thus, the quantum comb may be described as either an entangled state in the basis of individual frequencies or as a set of uncorrelated squeezed states in the supermode basis. As the individual supermodes are decoupled, it is straightforward to describe the effect of the cavity. Since the cavity does not spectrally filter the optical state, each eigenvector is resonant within the cavity, and a standard type-I OPO calculation is applied to each mode as a means to infer the output state. It follows from \cite{patera2010} that at the cavity threshold and zero Fourier frequency, the noise of the squeezed quadrature normalized to vacuum is given by
\begin{equation}\label{sqz-level}
V_k = \left(\frac{\Lambda_0-|\Lambda_k|}{\Lambda_0+|\Lambda_k|}\right)^2.
\end{equation}
Assuming a Gaussian shape for the coupling matrix $L_{m,n}$, these eigenvalues may be written as \cite{patera2010}:
\be
\label{eq:phases}
\Lambda_k = \Lambda_0 \, \rho^k
\ee
with
\begin{eqnarray}
\Lambda_0 &=& \pi^{\fr{1}{4}} \sqrt{\fr{2}{  \tau_\textrm{p} \, \omega_{\textrm{FSR}}}} \cdot \sqrt{\fr{\tau_\textrm{p}^2}{\tau_1^2 + \tau_\textrm{p}^2}} \, , \nonumber \\
\rho &=& -1 + 2 \sqrt{\fr{\tau_2^2}{\tau_1^2 + \tau_\textrm{p}^2}}, 
\end{eqnarray}
where $\tau_1= | k_\textrm{p}' - k_\textrm{s}'| l/\sqrt{10}, \, \tau_2 = \sqrt{|k_\textrm{s}'' | l } /(4 \sqrt{3})$, and $\tau_\textrm{p}$ is the temporal duration of the pump pulse. The nonlinear crystal length is specified by $l$ while $k'$ and $k''$ are the first and second derivatives, respectively, of the frequency-dependent wave vector for the pump (p) and signal (s) pulses. For realistic experimental parameters $\rho \simeq -1$ \cite{patera2010}. Hence, Eq.~\ref{eq:phases} corresponds to an alternating geometric progression of ratio $\rho$ whose first element $\Lambda_0$ is positive. 

The quadrature in which the $k^{\textrm{th}}$-mode of the field exhibits squeezing is determined by the phase of the corresponding eigenvalue $\Lambda_{k}$ \cite{patera2010}. As a result, the squeezing quadrature is predicted to alternate between the $x$ and $p$ quadratures with increasing mode index $k$. This theoretical prediction is well-verified in our experiment.

\subsection{Cluster State Basis}
\label{sse:clu_state_basis}

The output state of the SPOPO may be analyzed in a variety of different mode bases, and each basis reveals a specific entanglement structure \cite{Braunstein2005}. One class of entangled states of particular relevance for quantum information processing is that of cluster states. A cluster state is a highly entangled multimode state associated with a graph \cite{menicucci2011graphical}. This graph contains nodes that represent the various modes of the cluster state. An adjacency matrix $V$, which is real and symmetric, describes this graph and summarizes the entanglement connections among the various nodes (see Fig.~\ref{fig:clusters} for concrete examples). 
It has been shown that the cluster state defined by the adjacency matrix $V$ may be constructed from a set of independently $p$-squeezed input modes by combining them with a linear optical network in the appropriate manner \cite{vanLoock2007}. The action of this optical network can be mathematically described by a unitary matrix $U_V$ that transforms the collection of $N$ uncoupled $p$-squeezed modes into a $N$-mode cluster state. The mathematical relation between $V$ and $U_V$ is detailed below. 
\ffig{fig:clusters}{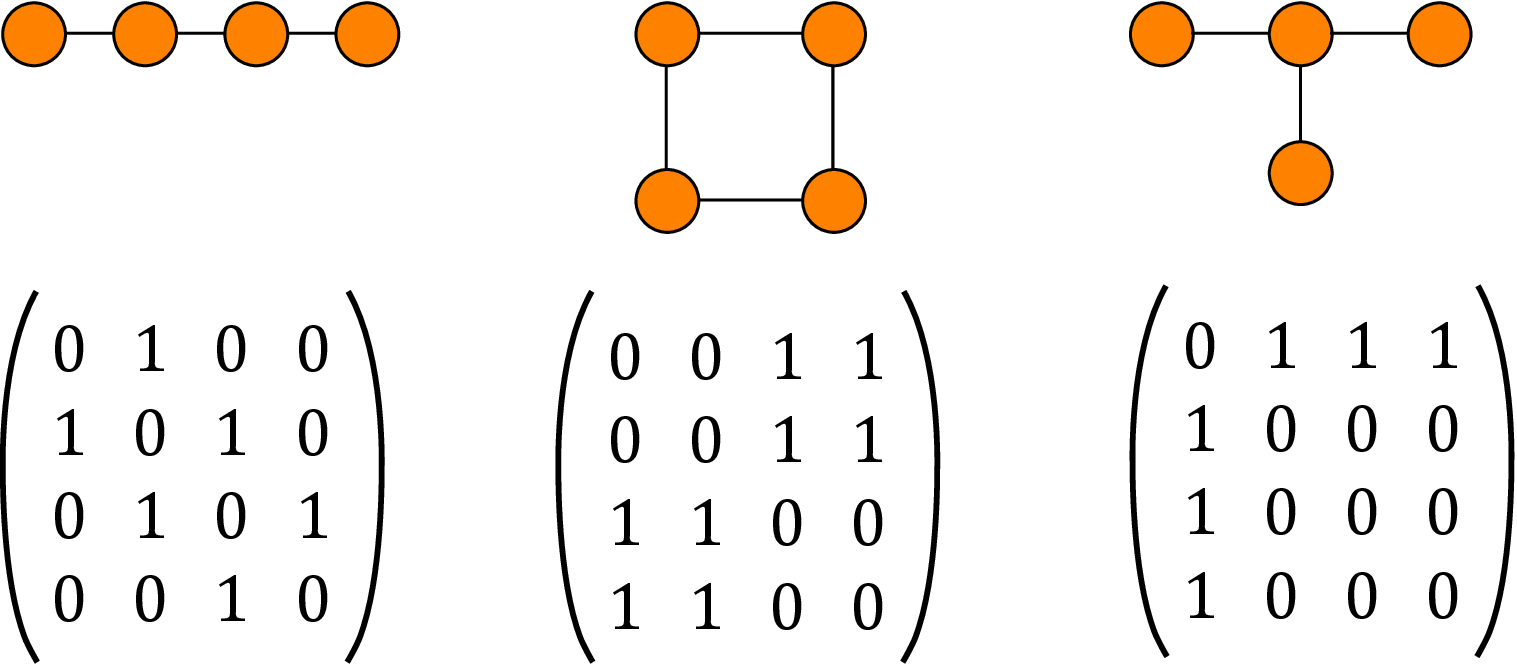}{Four-mode linear, square and T-cluster states (graphs and respective adjacency matrices $V_{\text{lin}}$, $V_{\text{square}}$, $V_{\text{T}}$).}{65mm}

The nullifier operators of a $N$-mode cluster state are derived from the adjacency matrix $V$ and may be written as:

\begin{equation} \label{eq:nullifier}
\hat{\delta}_i =  \left( \hat{p}_{i}^{C} - \sum_{j} V_{ij} \cdot \hat{x}_{j}^{C} \right),
\end{equation}
where $\hat x_i^C$ and $\hat p_i^C$ are the quadrature operators for the node $\hat a_i^C $, defined such that $\hat a_i^C  = \hat x_i^C  + \textrm{i}  \hat p_i^C $, and $i,j = 1,...,N$. Theoretically, a state is considered a cluster state of the adjacency matrix $V$ if and only if the variance of each nullifier approaches zero as the squeezing of the input modes approach infinity. From this definition, a unitary matrix $U_V$ may be constructed that defines the optical network for constructing a given cluster graph \cite{vanLoock2007}.

In order to determine the class of unitary matrices corresponding to a given adjacency matrix $V$, the unitary is decomposed as $U_{V} = X_{V} + \textrm{i} Y_{V}$, where $X_{V} = \textrm{Re} \left[ U_{V} \right]$ and $Y_{V} = \textrm{Im} \left[ U_{V} \right]$. The requirement that the variances of the nullifiers approach zero as squeezing goes to infinity is satisfied given the relation \cite{vanLoock2007}:
\be  \label{eq:LinNetwork}
Y_{V} = V \, X_{V}.
\ee
After exploiting the fact that $U_V$ is a unitary matrix (i.e., $X_VX^{T}_V+Y_VY^{T}_V = 1$ \cite{dutta1995real}), an initial unitary matrix $U_{V}^0$ is found for the desired graph state.

Importantly, the unitary matrix $U_{V}$ that creates a given cluster state is not unique, and the corresponding nullifier criteria of Eq.~\ref{eq:nullifier} is satisfied for a collection of different unitary matrices. In the case of finite squeezing, certain unitary matrices are more efficacious than others at creating the target cluster state (in the sense that they lead to a lower value of the nullifier variances). Other possible solutions may be obtained from the initial $U_{V}^0$ by multiplying it by a general, real orthogonal matrix $\mathcal{O}$ with $\mathcal{O} \mathcal{O}^T = \mathcal I$, i.e. $U'_{V} = U^0_{V} \mathcal{O}$  \cite{ferrini2013compact}. Given that $U^0_{V}$ forms a cluster state, it is straightforward to demonstrate that $U'_{V}$ also satisfies Eq.~\ref{eq:LinNetwork}. Thus, upon multiplying a specific $U_{V}$ by any orthogonal matrix, it is possible to span the complete space of physical unitary matrices satisfying Eq.~\ref{eq:LinNetwork}. 

In the case of the SPOPO, a large set of supermodes is available with each mode exhibiting a noise level given by Eq.~\ref{sqz-level}. In order to construct a cluster state from these modes, the $N$ modes displaying the highest degree of squeezing are selected, and the appropriate basis change defined by the matrix $U_{V}$ is applied. However, as the SPOPO output modes are not all squeezed along the same quadrature component, it is necessary to include an extra diagonal matrix $\Delta_{\text{sqz}} = \text{diag}\{e^{\textrm{i} \phi_1}, ..., e^{\textrm{i} \phi_N}\}$  that rotates each mode's squeezed quadrature into the common $\hat p$ direction. The transformation from the SPOPO squeezed modes to the desired cluster modes is then written as
\begin{equation} \label{eq:LinNetworkFourier}
\vec{a} \, ^{C} =  U_{V} \,  \Delta_{\text{sqz}} \, \vec{S} \, ,
\end{equation}
where $\vec{a} \, ^{C} = (\hat{a}_{1} ^{C}, ...,\hat{a}_{N} ^{C})$ is the collection of mode operators corresponding to each cluster node and $\vec S = (\hat S_1,...,\hat S_N)$ is the set of the leading $N$ supermodes as defined in Eq.~\ref{sqz-hamiltonian}. The remaining supermodes are left unchanged by the transformation and are not relevant for the $N$-mode cluster state considered here. In the present circumstance, a basis change $U_V$ is equivalent to a specific choice of measurement basis, which will be utilized to reveal cluster correlations embedded in the optical comb structure.

\section{Experimental Methods} \label{sec:methods}

The laser source is a titanium-sapphire mode-locked oscillator delivering $\sim 140 \textrm{fs}$ pulses ($\sim 6 \textrm{nm}$ FWHM) centered at 795nm with a repetition rate
of 76MHz. This source is frequency doubled in a 0.2mm BIBO crystal (single pass), and the resultant second harmonic pumps an OPO, which consists of a $2 \textrm{mm}$ BIBO crystal contained within a $\sim 4 \textrm{m}$ ring cavity exhibiting a finesse of $\sim 27$. The length of the cavity is locked to the inter-pulse spacing by injecting a phase-modulated near-infrared beam in a direction counter-propagating to the pump and seed. This locking beam is phase-modulated at 1.7MHz with an electro-optic modulator (EOM), and locking of the cavity length is accomplished with a Pound-Drever-Hall strategy. The cavity is operated below-threshold and in an unseeded configuration. Frequency correlations of the vacuum output are investigated with homodyne detection in which the local oscillator (LO) pulse form is manipulated with ultrafast pulse shaping methodologies. 

\ffigDouble{fig-experiment}{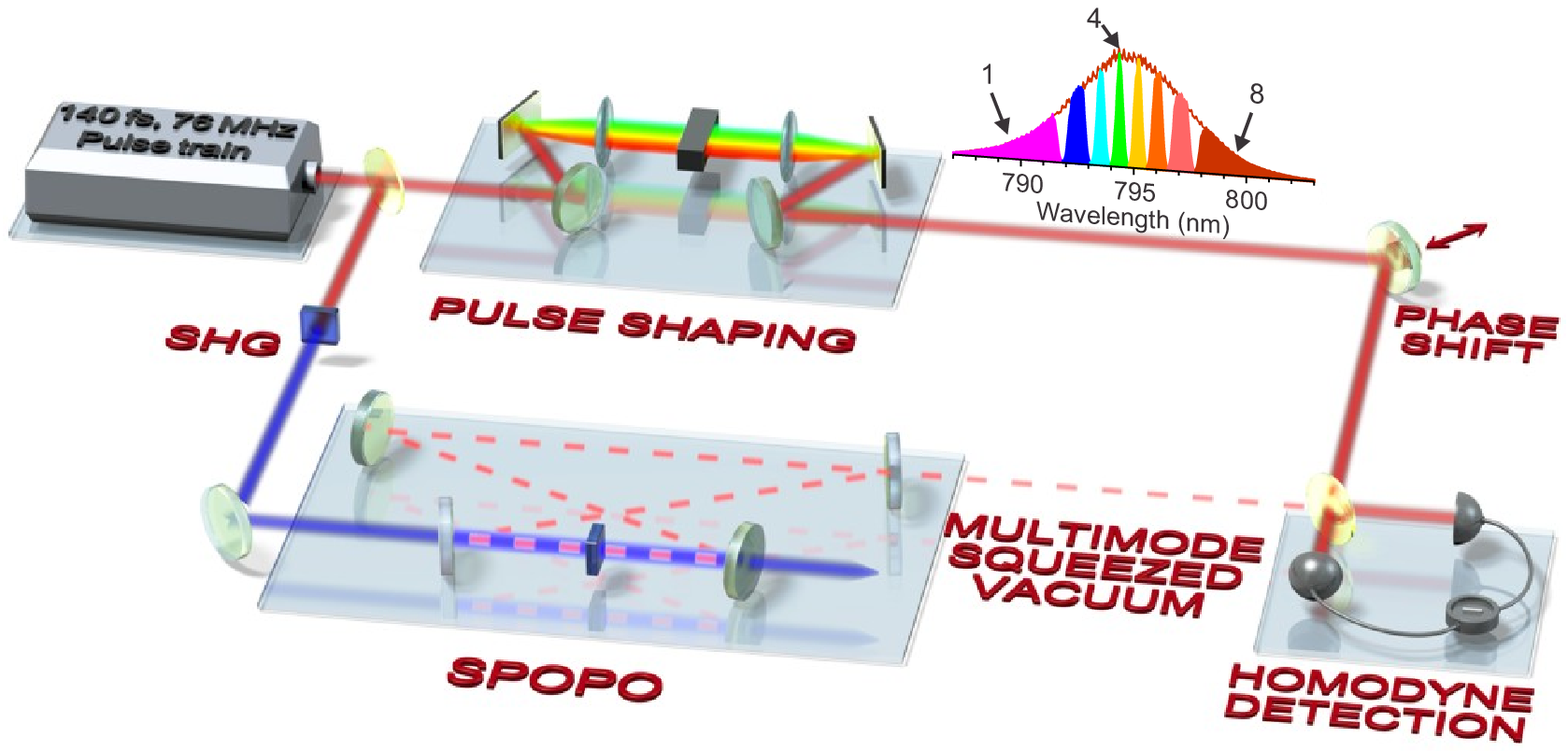}{Experimental
layout for the creation and characterization of multimode
frequency combs. A titanium-sapphire oscillator produces a $76
\textrm{MHz}$ train of $\sim 140 \textrm{fs}$ pulses centered at
795nm. Its second harmonic synchronously pumps an OPO. The cavity output is analyzed with
homodyne detection, where the spectral amplitude and phase of the local
oscillator (LO) are shaped. The LO shaper is depicted here in a transmissive geometry for clarity. By varying the relative phase between the shaped LO and the SPOPO output, the $x$- and $p$-quadrature noises of the quantum state projected onto the LO mode are measured.}{115mm}

A 4f-configuration shaper is constructed in a reflective geometry with a programmable 512 x 512-element liquid-crystal modulator in the Fourier plane. Application of a periodic spatial grating to the spatial light modulator induces diffraction of the spectrally-dispersed light. The amplitude and phase of the diffracted spectrum are independently controlled by the groove depth and position of the spatial grating, respectively \cite{nelson2005}. By varying the relative phase between the
shaped LO and the SPOPO output, a measurement is obtained of the $x$- and $p$-quadrature noises for the quantum state projected onto the spectral form of the LO mode (see Fig. \ref{fig-experiment}).

Light detection is performed with silicon photodiodes ($\sim 90\%$ detection efficiency, 100MHz detection bandwidth), and the homodyne visibility is $92\%$. The noise level of sidebands situated 1MHz from the optical carrier is then examined. The cumulative loss of the system is taken to be $\sim 25\%$, and the measured signals are corrected accordingly. The SPOPO generates vacuum squeezed at a level of $\sim 6 \textrm{dB}$ (corrected) when projected onto a local oscillator pulse taken directly from the titanium-sapphire laser. 

The noise properties of a Gaussian state are fully characterized in terms of its phase-space covariance matrix \cite{Braunstein2005}. This matrix of second-moments is directly reconstructed in the spectral domain by using the pulse shaper to measure noise correlations amongst different spectral regions. The LO spectrum  is divided into discrete bands of equal energy (e.g., in eight bands), and the amplitude and phase of each band may be individually addressed. Gaps between the individual spectral regions are intentionally imposed in order to ensure orthogonality of the different regions. Importantly, the supplemental loss incurred from the inclusions of these holes is not accounted for when correcting the noise levels. The $x$ quadrature is defined as the field quadrature of lowest noise for the unshaped LO pulse. The noise content of both the $x$- and $p$-quadratures for each spectral region and all possible pairs of regions are measured, which amounts to 36 measurements in the case of eight frequency zones. Individual covariance elements are then constructed according to the following relation:

\begin{eqnarray}
\langle x_{i} x_{j} \rangle &=& 
\left[ \langle (x_{i} + x_{j})^2 \rangle - \frac{P_{i}}{P_{i}+P_{j}} \langle x_{i}^2 \rangle - \frac{P_{j}}{P_{i}+P_{j}} \langle x_{j}^2 \rangle \right] \nonumber \\ && \times \, \frac{P_{i}+P_{j}}{2 \sqrt{P_{i} P_{j}}},
\end{eqnarray}
where $P_{i}$ and $P_{j}$ are the optical powers of frequency bands $i$ and $j$, respectively, which are measured with the homodyne photodiodes. 

Importantly, it has been verified that the LO phase dependence for each of the 36 noise measurements follows the same dependence as that of the unshaped LO reference. Consequently, the lowest noise level for every spectral combination is present in the $x$ quadrature, i.e., there is no rotation of the squeezing ellipse between successive measurements. Additionally, it has been observed that cross-correlations of the form $\langle x \, p \rangle$ are absent, which permits the covariance matrix to be cast in a block diagonal form: one block for the $x$-quadrature and one block for the $p$-quadrature.

We have seen that a good reconstruction of the quantum comb is accomplished with eight discrete LO spectral bands (or with ten bands as presented in \cite{Roslund2013}). However, it is feasible to perform the measurements with a reduced number of spectral regions depending upon the application. In what follows, results will be presented for a variety of different dimensionalities of the LO frequency space. 

\section{Experimental Results} \label{sec:results}

\subsection{State Reconstruction}

\ffig{fig-matrices}{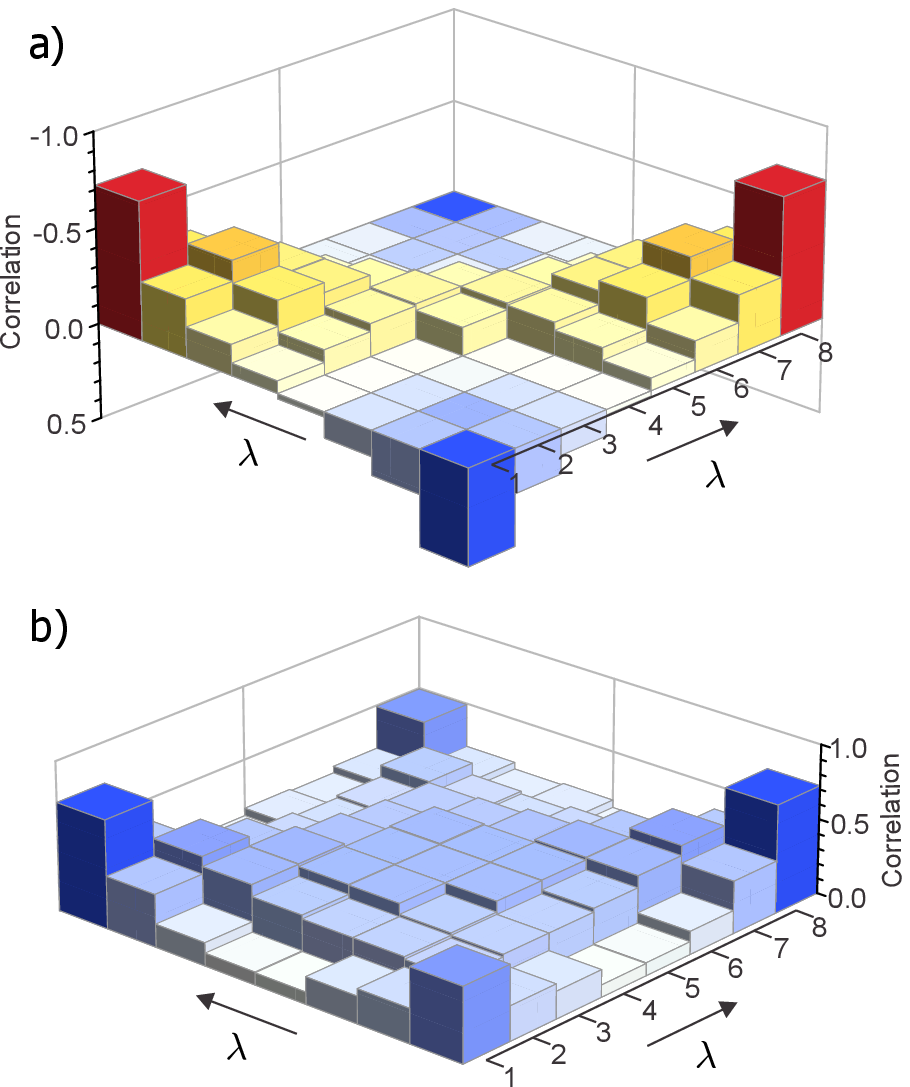}{Experimentally measured quantum noise matrices for the $x$- (a) and $p$- (b) quadratures. The noise correlation matrix is defined as in Eq.~\ref{eq:corr_matrix}. Each matrix reveals significant correlations among the frequency bands of the comb.}{85mm}

The full covariance matrix of the quantum comb is reconstructed following the 36 requisite homodyne measurements. Fluctuations and correlations departing from the vacuum level are depicted with the noise correlation matrix, which is defined as: 

\begin{equation}
C^{x}_{i,j} = \langle x_{i} x_{j} \rangle / \sqrt{\langle x_{i}^2 \rangle \langle x_{j}^2 \rangle} - \delta_{i,j} \langle x_\textrm{vacuum}^2 \rangle / \langle x_{i}^2 \rangle
\label{eq:corr_matrix}
\end{equation}
for the $x$-quadrature with a similar definition for the $p$-quadrature. The retrieved correlation matrices for the two field quadratures are shown in Fig.~\ref{fig-matrices}. The spectral wings of the state's $x$-quadrature possess excess noise as compared to the frequency bands near the central wavelength; however, the strongest correlations are also evident in the wings. Qualitatively, this situation is consistent with a two-mode squeezed state, in which tracing out a single mode results in a thermal state (i.e., quadrature-independent excess noise). 

Entanglement among various frequency bands is quantitatively assessed with the positive partial transpose (PPT) criterion for continuous variables \cite{simon2000}, which probes the inseparability of a given state bipartition. A bipartition is created by dividing the eight frequency bands of the comb into two sets. The transposition of one of these sets is achieved through a sign change of all momenta operators $\hat{p}_{i}$ contained within the set: $\left( \hat{x}_{i}, \hat{p}_{i} \right) \rightarrow \Gamma_{ii} \cdot \left( \hat{x}_{i}, \hat{p}_{i} \right) = \left( \hat{x}_{i}, -\hat{p}_{i} \right) $. This time-reversal operation creates a new covariance matrix $V_{\textrm{PPT}} = \Gamma V \Gamma$, which must continue to satisfy the Heisenberg uncertainty relation: $P = \Gamma V \Gamma - i \Lambda \geq 0$, where $\Lambda$ is the symplectic matrix \cite{Braunstein2005}. The two bipartitions are entangled if the Heisenberg matrix $P$ is not positive definite. Given eight distinct spectral bands, 127 unique frequency band bipartitions exist. Each of these possible bipartitions is subjected to the PPT criterion, and the minimum eigenvalue of $P$ is shown in Fig.~\ref{fig-ppt}. As seen in the figure, every possible state bipartition is entangled. The absence of any partially separable form implies that the SPOPO output constitutes a genuine 8-partite state in which each resolvable frequency element is entangled with every other component \cite{Braunstein2005}. Accordingly, the downconversion of a femtosecond frequency comb indeed creates a quantum object exhibiting wavelength entanglement that extends throughout the entirety of its structure. 
 
Two distinct bands of PPT values are evident in Fig.~\ref{fig-ppt}. The band exhibiting a higher degree of entanglement (lower PPT value) is composed of all bipartitions that separate the highest and lowest frequency zones (pixels 1 and 8). Within this band, the most strongly entangled form results from division of the spectrum at the central wavelength. Conversely, the alternative band (higher PPT value) consists of those partitions in which these extreme spectral zones are not disconnected. The partition that dissociates the two spectral wings from the remaining spectrum corresponds to the most weakly entangled structure. Consequently, the spectral wings may be considered as reproducing the situation of two-mode entanglement, which is consistent with the structure of the covariance matrix. 

However, the multimode character of the comb can not be directly inferred from the high degree of multipartite entanglement. It is well known that the bipartition of a single mode squeezed field creates two entangled modes that satisfy the PPT criteria. As a means for comparison, Fig.~\ref{fig-ppt} also includes the same 127 spectral partitions for a simulated single mode field with quadrature values that correspond to those of the first comb supermode. As seen in the figure, all of these bipartitions also satisfy the inseparability criterion. The minimum eigenvalue of $P$ no longer depends upon the symmetry of the bipartition but only upon the relative power between the two partitions. Nonetheless, PPT values for the single mode case are weaker than those observed for the comb, which provides a first indication of the comb's multimode character.

\ffig{fig-ppt}{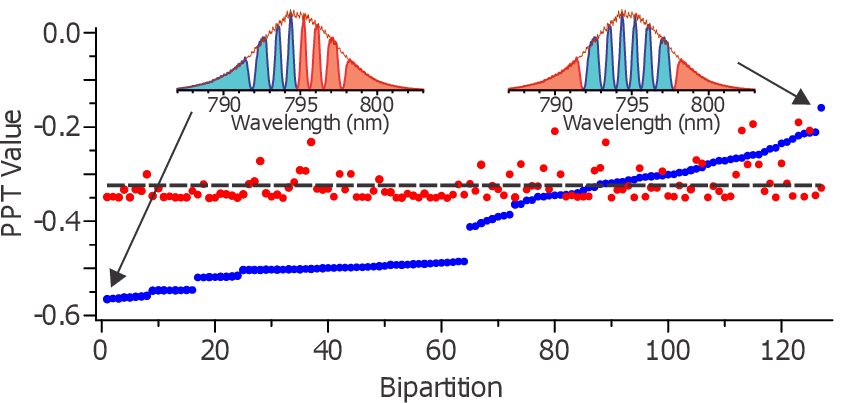}{The PPT (blue) inseparability criteria for all 127 bipartite combinations of the 8 spectral bands. All 127 bipartitions possess a PPT value below the entanglement boundary of 0.0, which indicates complete non-separability for the state. The PPT criteria is also applied to a simulated single mode squeezed state (red) with noise parameters corresponding to the first supermode. The single mode PPT values are ordered according to the full PPT values. The black dotted line represents the mean single mode PPT value. All full PPT bipartitions below this line are indicative of multimode character. }{85mm}

\subsection{Eigenmode Decomposition}

Although multipartite frequency entanglement is relevant for the creation of specialized entangled states, it is an extrinsic property of the comb. For example, as explained above, the PPT criteria depends upon a predefined allocation of individual frequency bands. Multipartite character may always be imparted to a single mode quantum object by simply dividing it with a beamsplitter. 

However, the basis change introduced in Eq.~\ref{sqz-hamiltonian} may be implemented as a means to recover a set of independently squeezed spectral modes embedded in the beam. This generalized Schmidt decomposition is achieved by diagonalizing the recovered covariance matrix to reveal a set of decorrelated supermodes $\hat{S}_{k}$. When the matrices of Fig.~\ref{fig-matrices} are eigendecomposed, it is observed that although the individual $x$ and $p$ block eigenvectors are quite similar, they are not exactly equal. This implies that a common mode basis is not able to simultaneously diagonalize the two quadrature blocks.

In order to understand the physical origin of this effect, the complete decomposition of the symplectic matrix responsible for creating the multimode state is considered. The Bloch-Messiah reduction \cite{Braunstein2005-irreducible, dutta1995real} allows any symplectic transformation to be decomposed into an initial basis change, a perfect multimode squeezer, and a final basis change. When the input state to this transformation is vacuum, the first basis rotation is arbitrary, and the resultant multimode state may be understood as an assembly of squeezers in a given eigenbasis (as seen in Eq.~\ref{sqz-hamiltonian}). However, when the input state either contains classical noise or is not pure, both of these basis rotations become meaningful. 

Application of the Bloch-Messiah reduction to a covariance matrix reveals the Williamson (or ``symplectic'') eigenvalues as well as the mode structures for both the classical noise and quantum squeezers. It is these Williamson eigenvalues that indicate the existence of residual classical noise on the input state. Importantly, in the presence of excess classical noise, the quantum squeezer basis and the supermode basis do not necessarily correspond. In the present experiment, the input state to the cavity is vacuum, which implies that residual classical noise is introduced by loss mechanisms. Correspondingly, the fact that the $x$ and $p$ blocks of the covariance matrix are not diagonalized by a common basis indicates that the loss mechanism is spectrally dependent (e.g., non-uniform transmission profile of the SPOPO output coupler).

A Bloch-Messiah reduction of the 8-mode covariance matrix was implemented in order to reveal the full structure of the comb state. The Williamson eigenvalues possess values close to unity, which indicate that the purity of the comb state is quite high.  Additionally, the bases of the classical noise eigenmodes and the squeezed modes are independently uncovered. The squeezed mode basis remains largely unchanged from run to run, while the basis associated with the classical noise exhibits a large degree of variation that depends upon specific experimental conditions. This effect arises because the classical noise is relatively small compared to the quantum properties of the comb, and the eigenvalues are nearly degenerate. Consequently, the extraction of well-defined supermodes from the experimental covariance matrix is feasible even though the matrix can not be placed in a perfectly diagonal form due to the influence of classical noise. 

In practice, the experimental supermodes are recovered with a more pragmatic strategy. Upon eigendecomposition of the covariance matrix, the modes exhibiting squeezing are observed to alternate between the $x$ and $p$ quadratures. The eigenstructures corresponding to the anti-squeezed modes, which likewise alternate between the two quadratures, 
exhibit increased robustness to noise (which arises from their increased angular contribution to the squeezing ellipse). In order to determine the covariance matrix of an entirely decoupled mode set, the eight anti-squeezed eigenmodes are orthogonalized with a Gram-Schmidt procedure, and the covariance matrix is re-expressed in terms of this newly orthogonal basis. The resulting matrix is nearly diagonal and contains the squeezing value for each orthogonalized mode on its diagonal. 

When performing the individual frequency band measurements utilized to construct the covariance matrix, multiple oscillations of each noise trace are collected in order to estimate the uncertainty of the corresponding squeezing and anti-squeezing levels. These uncertainties are exploited to assess the error level of each supermode squeezing value with a stochastic sampling methodology. Noise values for a particular spectral band combination are drawn from a normal distribution with a mean specified by the average of all identified peaks or valleys and a variance given by the variance of the extrema. A collection of $10^{4}$ individual covariance matrices is amassed, where each matrix is assembled by drawing samples from the necessary normal distributions. The Gram-Schmidt orthogonalization procedure is implemented for each matrix, which yields a squeezing spectrum and mode set. The mean squeezing spectrum is shown in Fig.~\ref{fig-eigenvalues} for the situations of 4, 6, and 8 discrete spectral regions. 

In each case, the mean spectrum is generally noise-robust. For the leading modes, a larger overall squeezing level is observed for a smaller number of pixels, which is a consequence of the imposed spectral gaps. However, when the covariance matrix is reconstructed with a larger number of pixels, the eigenspectrum exhibits more modes. An increase in the number of available pixels is needed to replicate the spectral complexity of higher-order supermodes. In the case of 8 unique frequency bands, up to 7 squeezed modes are contained within the conglomerate comb structure (while 8 squeezed modes were found in the ten-pixel spectral reconstruction performed in \cite{Roslund2013}). The quadrature in which each of these modes exhibits noise reduction ($x$ or $p$) alternates between successive modes in agreement with theoretical predictions \cite{patera2010}. As such, the SPOPO behaves as an \emph{in situ} optical device consisting of an assembly of independent squeezers and phase shifters. 

\ffig{fig-eigenvalues}{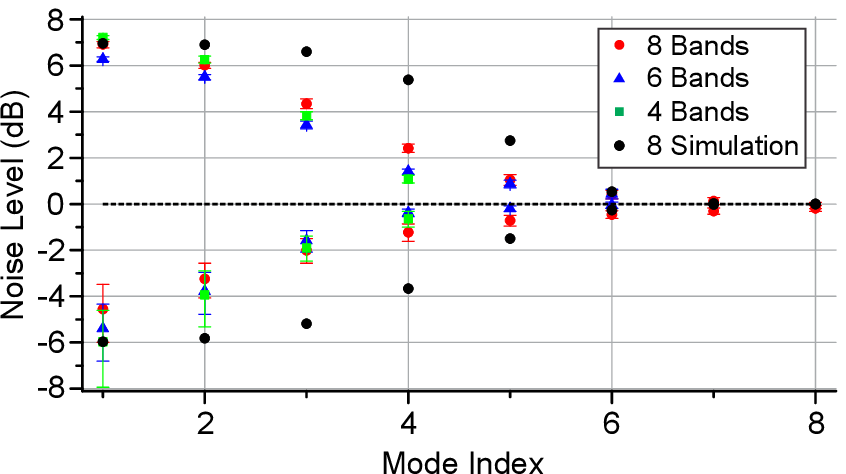}{Mean noise levels and uncertainties (dB) for each of the orthogonalized Gram-Schmidt modes. The mean eigenspectra are shown for 8 (red), 6 (blue), and 4 (green) unique frequency bands. The simulated eigenvalues corresponding to 8 frequency bands are shown for comparison (black).}{80mm}

The orthogonalized modes that originate from covariance matrices comprised of four and eight spectral zones are shown in Figs.~\ref{fig-supermodes}a and \ref{fig-8modes}, respectively. The spectral makeup of each retrieved experimental mode displayed in Fig.~\ref{fig-supermodes}a follows the form of a Hermite-Gauss polynomial, which approximates the predicted supermode profile \cite{patera2010}. However, as mentioned above, it becomes evident that the spectral complexity of higher-order supermodes is only reproducible with an increase in the number of pixels. Additionally, the spectral width $\Delta \lambda_{k}$ of supermodes following a Hermite-Gauss progression increases with the mode index $k$ as $\Delta \lambda_{k} =\sqrt{2 k+1} \cdot \Delta \lambda_{0}$. 

\ffig{fig-supermodes}{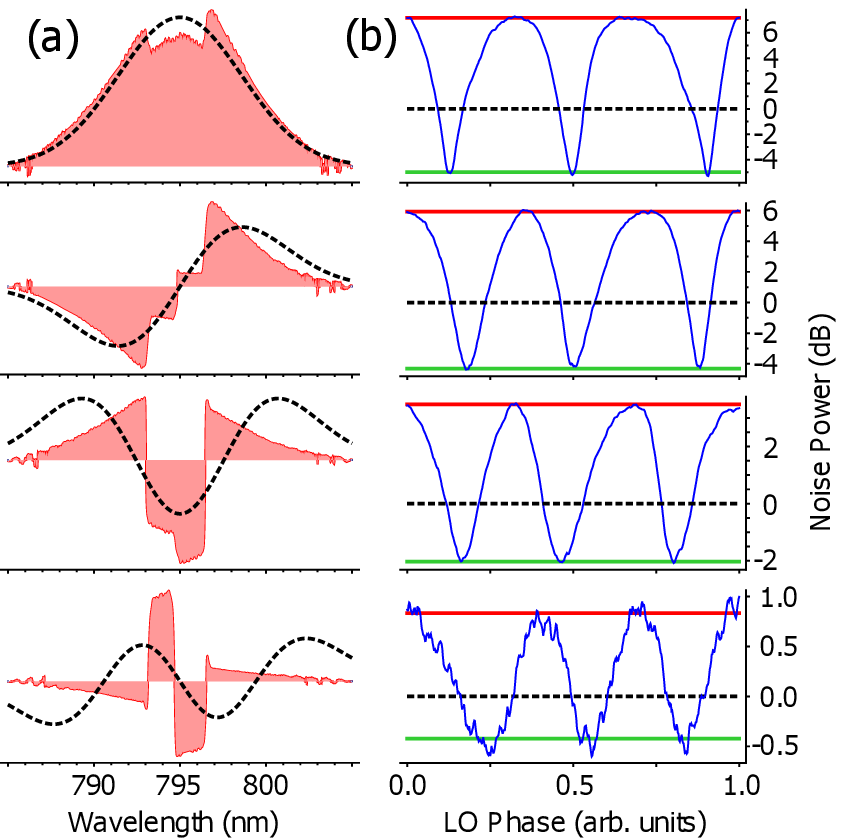}{(a) Retrieved experimental supermodes with the spectral gaps removed. The field of each supermode is measured with spectral interferometry. (b) Noise traces corresponding to each of the experimental supermodes.}{85mm}

In order to assess the impact of both the LO bandwidth and the number of independent shaper elements on the observed squeezing levels, a series of simulations was performed utilizing the current experimental parameters. These simulations were performed by directly calculating the supermodes from the phase-matching properties of a BIBO crystal \cite{patera2010} while assuming a perfect cavity with a bandwidth of 50nm. The purity of the state is taken to match the quadrature noises of the first supermode. With these parameters, the cavity output contains $\sim 25$ modes with an equivalent level of squeezing. 

Subsequently, a 8-frequency pixel homodyne detection apparatus is simulated without accounting for the supplemental losses incurred by the gaps between pulse shape pixels. The resulting eigenspectrum is shown in Fig.~\ref{fig-eigenvalues}. The simulation results are consistent with the fact that the spectral overlap diminishes between the fixed bandwidth of the LO spectrum and each progressively broadened supermode. This decline in the spectral overlap becomes especially prominent in the wings. While $\sim 25$ modes are initially present in the comb state, only $\sim 5$ are detected as seen in Fig.~\ref{fig-eigenvalues}. Two technical effects account for this loss of modes in the detection process. First, the fixed bandwidth of the LO only achieves perfect overlap with the first supermode. In addition, the high spatial frequency of the spectral structures created by the pulse shaper begin to exhibit appreciable spectral overlap with very high-order supermodes that are not squeezed. Both of these limitations introduce vacuum into the measurement, which degrades the overall squeezing levels. Consequently,  the current observation of 7 squeezed modes does not represent an inherent upper limit to the quantum dimensionality of comb states. With the use of broader bandwidth LO pulses, increased spectral resolution, and large cavity bandwidths (all achievable experimentally), states possessing as many as  $\sim 100$ squeezed modes are expected \cite{patera2010}.

\ffig{fig-8modes}{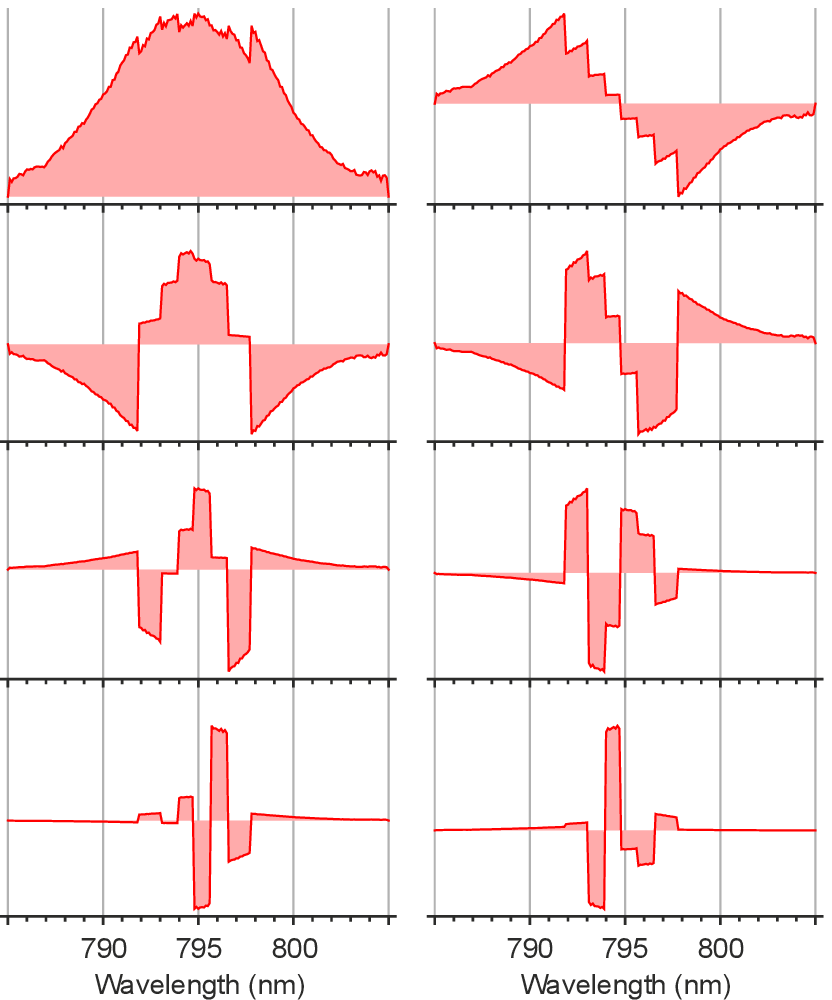}{Amplitude spectra corresponding to each of the orthogonal supermodes retrieved from the covariance matrix shown in Fig.~\ref{fig-matrices}.}{80mm}

\subsection{Eigenmode Corroboration}

The supermodes displayed in Fig.~\ref{fig-supermodes}a constitute an uncoupled set of independent squeezed states that can serve as a resource for the construction of specialized entangled structures, such as cluster states. As such, it is important to validate the structure and squeezing level of the modes retrieved from the covariance matrix. 

Each of the modes derived from the covariance matrix is written directly onto the pulse shaper after bridging the spectral gaps that were imposed in the frequency band basis. The corresponding noise traces are seen in Fig.~\ref{fig-supermodes}b. Most importantly, each of these four modes exhibits squeezing at a level in accordance with that retrieved from the covariance matrix. Furthermore, the quadrature of squeezing alternates between successive modes. Thus, when the sum of the first two modes is written to the shaper, excess noise is present in both quadratures (not shown). Consequently, the SPOPO simultaneously generates states that are squeezed in either the amplitude or the phase quadrature. 

\subsection{Discussion on State Purity}

The purity $\mathcal{P}$, which is an intrinsic property of the state, is accessible from the covariance 
matrix with the relation $\mathcal{P} = 1 / \sqrt{\textrm{det}(\Sigma^E)}$, where $\Sigma^E$ is 
the measured covariance matrix. The covariance matrix eigenvalues shown in Fig.~\ref{fig-eigenvalues} enable comparison of the state purity for 4-, 6- and 8-spectral-zone divisions of the LO spectrum. The purity values were also measured for different pump powers (not shown), and the observed variation follows the expected behavior for the output state of an OPO. 

The imposition of gaps between the discrete spectral regions represents a loss, and therefore decreases the state purity. As fewer gaps are necessary to create four spectral zones, a higher state purity is expected for the four-band matrix as compared to the eight-band matrix. As seen in Fig.~\ref{fig-eigenvalues}, although the overall squeezing levels are made similar with an appropriate tuning of the pump power, the four-band state possess a slightly higher purity. By fine adjustment of the experimental parameters, a global purity ranging from $\mathcal{P} \sim 0.7 - 0.8$ is achievable while maintaining significant squeezing levels. The ability to achieve high purity states while maintaining their multimode nature constitutes an important resource for the construction of network structures. 

\section{Cluster State Analysis} \label{sec:clusters}
 \subsection{Creation of the Cluster Basis}

In the previous section, a basis change of the covariance matrix allowed retrieval of the theoretically predicted supermodes. Similarly, an analogous procedure may be applied to construct cluster state bases as defined by Eq.~\ref{eq:LinNetworkFourier}. In doing so, the feasibility for creating cluster states from experimentally retrieved covariance matrices may be directly probed. 

Application of the Gram-Schmidt orthogonalization method described above defines a rotation matrix $U_{T}$, which transforms the correlated pixel bands $\vec{a} \, ^{\text{pix}}$ into a set of nearly decorrelated experimental supermodes $\vec{S}$ through the relation $\vec{S} = U_{T}^{-1} \, \vec{a} \, ^{\text{pix}}$. 
The experimentally retrieved supermodes shown in Fig.~\ref{fig-supermodes} exhibit squeezing in alternating quadratures. Hence, it becomes necessary to apply a mode-selective phase rotation in order to transfer each mode's squeezing axis into the same direction. In order to accomplish this task, the phase-shift matrix $\Delta_{\text{sqz}} = \textrm{diag\{i,1,...,i,1\}}$ is applied. Subsequently, the cluster state corresponding to a particular adjacency matrix $V$ is constructed by applying the appropriate unitary matrix $U_{V}$. Thus, the total transformation relating the original pixel basis to the one parameterizing the cluster state is described by 

\begin{equation} 
\label{eq:utot}
\vec{a} \, ^{C} =  U_{V} \,  \Delta_{\text{sqz}} \, U_{T}^{-1} \, \vec{a} \, ^{\text{pix}} \equiv U_{\textrm{tot}} \, \vec{a} \, ^{\text{pix}} ,
\end{equation}
where $U_{V}$ satisfies Eq.~\ref{eq:LinNetwork}. 

Among the group of matrices $U_V$ that satisfy Eq.~\ref{eq:LinNetwork}, one is selected that minimizes the nullifier variances of Eq.~\ref{eq:nullifier} for the transformed cluster modes $\vec{a} \, ^{C}$. This is accomplished by parameterizing the most general orthogonal matrix in terms of a collection of angles: $\mathcal O (\vec{\theta})$. Upon doing so, all physically relevant cluster unitaries are spanned as $U_V = U_V^0 \, \mathcal O(\vec{\theta})$ in line with the discussion of Sec.~\ref{sse:clu_state_basis}. An evolutionary strategy \cite{roslund2009accelerated} is employed to search for a set of angles $\vec{\theta}$ that minimizes the nullifier variances in Eq.(\ref{eq:nullifier}) for the transformed cluster modes $\vec{a} \, ^{C}$.

A symplectic transformation $S_{\textrm{tot}}$ corresponding to the optimal unitary matrix $U_{\textrm{tot}}$ may then be written as
$S_{\textrm{tot}} = \left( \begin{array}{cccccccc}
X_{\textrm{tot}} & - Y_{\textrm{tot}} \\
Y_{\textrm{tot}} &  X_{\textrm{tot}} \\ \end{array} \right)$ with $U_{\textrm{tot}} = X_{\textrm{tot}} + \textrm{i} Y_{\textrm{tot}}$ \cite{dutta1995real}. This transformation is applied to the covariance matrix measured in the pixel basis $\Sigma^E$ in order to yield the covariance matrix of the cluster state:
\be
\label{eq:trasfa}
\Sigma^{C} = S_{\textrm{tot}}  \Sigma^{E} S_{\textrm{tot}}^{T}.
\ee 
Individual cluster correlations are then verified by determining whether the set of nullifier variances for each cluster state (as defined by Eq.~\ref{eq:nullifier}) lie below the shot noise level, i.e., $\delta_{i} < \delta_{\textrm{shot}} = 1$ for $i=1...N$. In this context, the shot noise level is defined as the nullifier variances obtained with a vacuum input to the linear network.
  
\subsection{Six-mode Cluster States}

In order to provide specific examples as to the potential for creating cluster states within the quantum comb, several different six-mode cluster structures are considered with corresponding graphs displayed in Table~\ref{tab:tablea}. 
The requisite squeezed input modes are those originating from a covariance matrix measured in a basis of six spectral zones. Following optimization of the orthogonal matrix $\mathcal O (\vec{\theta})$, the nullifier variances for each cluster structure are computed from the cluster covariance matrix $\Sigma^{C}$ as defined in Eq.~\ref{eq:trasfa}. Each set of variances is normalized to the respective shot noise levels. 

The cluster analysis is performed with two different sets of squeezed input modes. Set A exhibits relatively low input squeezing levels but a high state purity, while Set B displays high input squeezing levels and a lower state purity. This latter set of supermodes is obtained by operating closer to the cavity threshold. In both cases, each of the requisite nullifiers possesses a value below the shot noise level for all of the considered cluster structures. The higher input squeezing values present in Set B result in improved cluster correlations as highlighted by nullifier variances significantly below the shot noise limit. 


\begin{widetext}

\begin{table}
\centering
\begin{tabular}{|c|c|c|c|}
\cline{2-4}
\multicolumn{1}{c|}{}& Graph & Nullifiers $\{ \delta_i \}$ (Set A) & Nullifiers $\{ \delta_i \}$ (Set B) \\
\hline
Linear  &  
\includegraphics*[width=1cm]{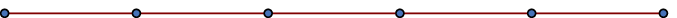}
&  \{0.85, 0.74, 0.69, 0.67, 0.75, 0.86\} &\{0.76, 0.58, 0.42, 0.46, 0.55, 0.78\}  \\
\hline
Hexagon & 
\includegraphics*[width=1cm]{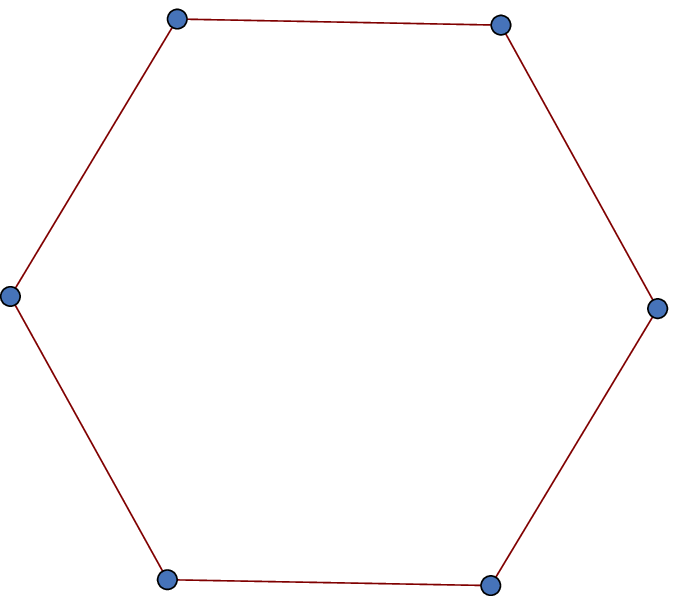}
& \{0.79, 0.76, 0.76, 0.73, 0.76, 0.70\} & \{0.55, 0.55, 0.59, 0.52, 0.59, 0.63\} \\
\hline
Connected Hexagon &
\includegraphics*[width=1cm]{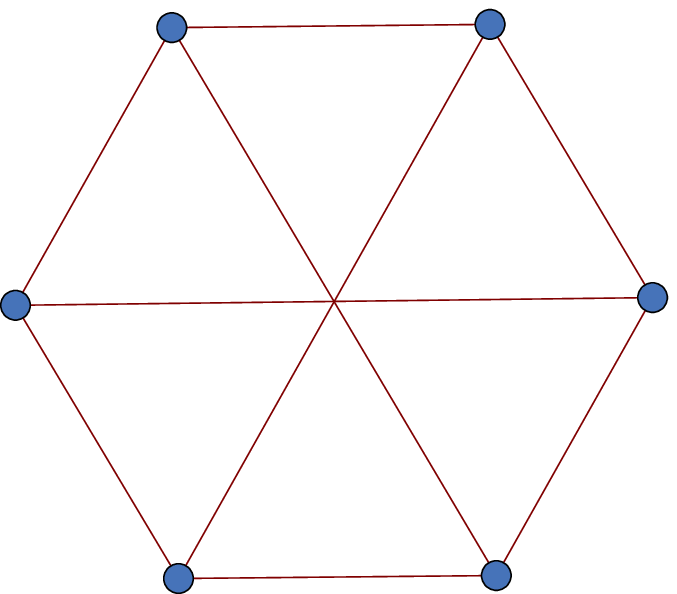}  
& \{0.66, 0.65, 0.67, 0.65, 0.66, 0.63\} & \{0.43, 0.37, 0.44, 0.36, 0.44, 0.33\} \\
\hline
Maximally Connected Hexagon & 
\includegraphics*[width=1cm]{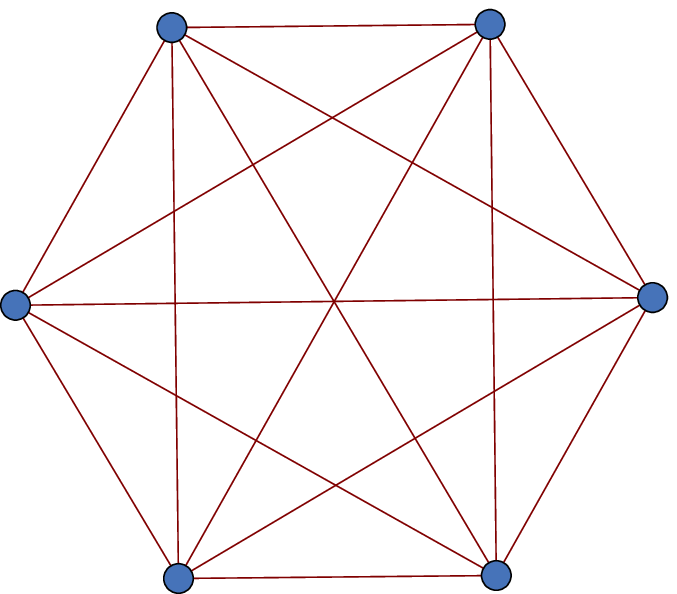}  
&   \{0.67, 0.68, 0.65, 0.67, 0.67, 0.64\} & \{0.42, 0.43, 0.41, 0.45, 0.37, 0.27\} \\
\hline
Prism & 
\includegraphics*[width=1cm]{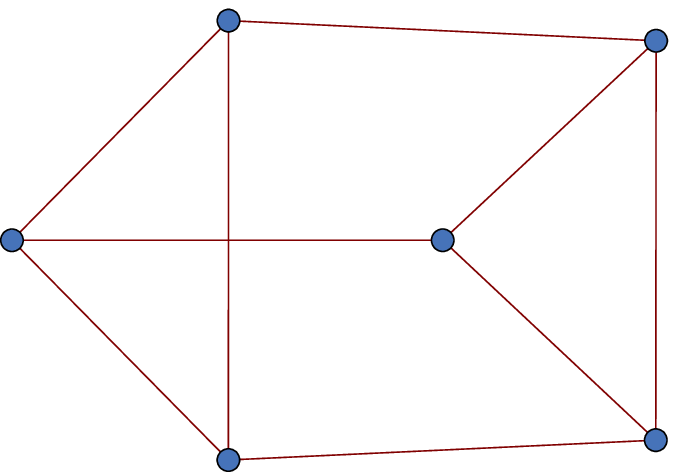}  
& \{0.66, 0.71, 0.73, 0.66, 0.71, 0.73\} & \{0.42, 0.47, 0.55, 0.42, 0.47, 0.55\} \\
\hline
Connected Square Pyramid & 
\includegraphics*[width=1cm]{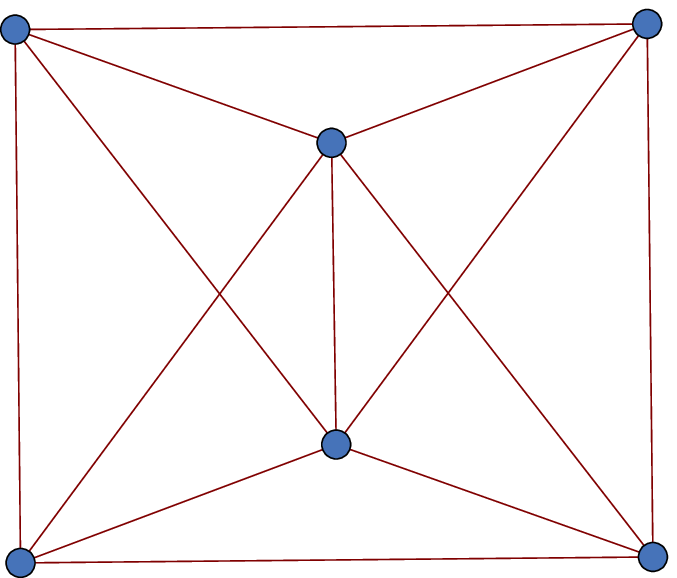}  
&  \{0.67, 0.64, 0.65, 0.64, 0.65, 0.67\} & \{0.41, 0.37, 0.37, 0.37, 0.37, 0.41\} \\
\hline
Double Square & 
\includegraphics*[width=1cm]{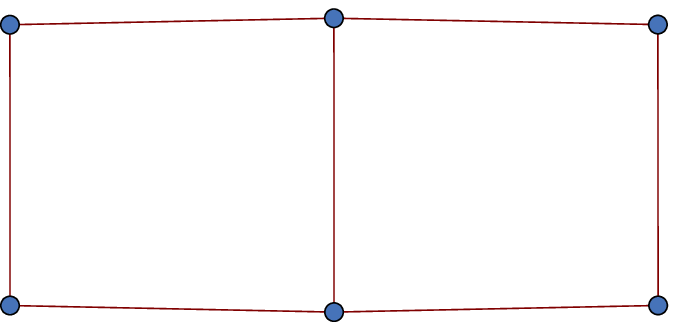}  
&  \{0.71, 0.75, 0.66, 0.65, 0.71, 0.75\}  & \{0.48, 0.60, 0.44, 0.36, 0.48, 0.60\} \\
\hline
Connected Double Square & 
\includegraphics*[width=1cm]{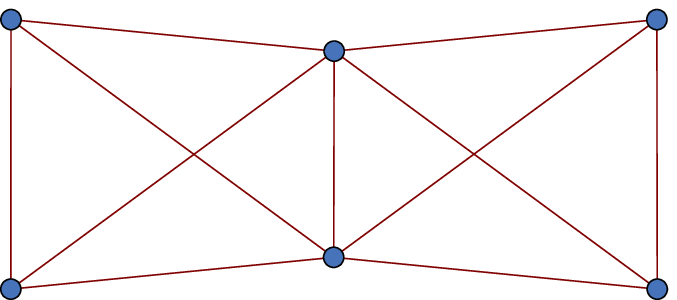}  
&  \{0.71, 0.71, 0.66, 0.66, 0.72, 0.71\} & \{0.50, 0.53, 0.36, 0.36, 0.49, 0.53\} \\
\hline
Pentagonal Pyramid & 
\includegraphics*[width=1cm]{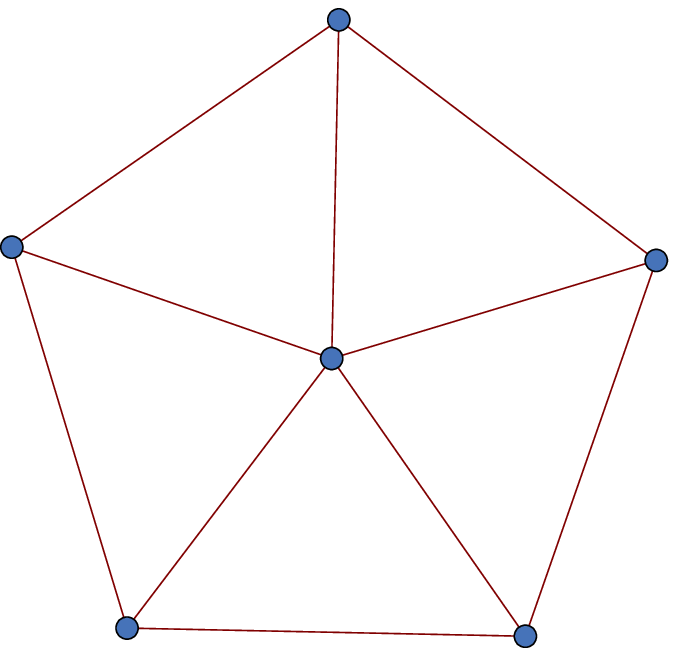}  
&  \{0.69, 0.73, 0.73, 0.70, 0.66, 0.71\}  & \{0.59, 0.48, 0.41, 0.48, 0.60, 0.40\} \\
\hline
\end{tabular}
\caption{Cluster state nullifiers $\{ \delta_i \}$ normalized with respect to the corresponding shot noise value of $\delta_{\textrm{shot}} = 1$. Set A consists of input modes with a maximum squeezing value of $-2.34~\textrm{dB}$ and a high state purity of $\mathcal{P} = 0.84$. Conversely, Set B utilizes input modes exhibiting a maximum squeezing level of $-6.48~\textrm{dB}$ and a state purity of $\mathcal{P} = 0.69$. In both cases, all of the considered cluster states are realized. The higher input squeezing levels associated with Set B result in enhanced violations of the nullifier variance criteria.\label{tab:tablea}}
\end{table} 

\end{widetext}

\section{Discussion} \label{sec:discussion}

The intrinsic entanglement of the quantum frequency comb provides an irreducible, universal quantum resource \cite{Braunstein2005-irreducible} of direct relevance for quantum information processing. From this multimode resource, the creation of cluster states or any user-defined structure is affected through an appropriate basis change. In particular, the frequency entanglement present in the comb is arranged in such a way that multiple cluster states are simultaneously embedded in its structure. Importantly, the realization of these states does not necessitate any change in the optical architecture itself, but rather simply in the manner by which the state is measured. The projective measurements necessary to realize such entangled states may be implemented with any variety of spectrally-resolved homodyne detection, including pulse shaping of the local oscillator. Theoretical analysis has proven that it is possible to fabricate the cluster structures necessary for computation from the modes contained within the quantum comb \cite{ferrini2013compact}. Likewise, the experimental feasibility of achieving basis transformations through measurement has already been demonstrated in the analogous domain of spatially multimode beams \cite{Armstrong2012}.

It is important to stress that the technical difficulties currently limiting the number of observed squeezed modes to $\sim 10$ are not fundamental to the methodology. Suitable improvements to the experimental setup (e.g., better adapting the pump spectrum, etc.) are expected to lift these obstacles. Accordingly, simulations predict that $\sim 100$ squeezed modes are expected to be embedded in the frequency structure of the quantum comb \cite{patera2010}. Such a resource is scalable and ideal for implementing quantum information protocols. 

In summary, the parametric downconversion of ultrafast frequency combs provides a practical and scalable multimode resource. The ability to generate top-down entanglement amongst thousands of frequencies with a single nonlinear interaction provides a unique capability and bodes well for the continued development of specialized quantum networks within highly multimode structures. 

\acknowledgments

This work is supported by the European Research Council starting grant Frecquam and the French National Research Agency project Comb. C.F. is a member of the Institut Universitaire de France. J.R. acknowledges support from the European Commission through Marie Curie Actions, and Y.C. recognizes the China Scholarship Council.

\bibliographystyle{apsrev}
\bibliography{multimode-BIB}

\begin{thebibliography}{32}
\expandafter\ifx\csname natexlab\endcsname\relax\def\natexlab#1{#1}\fi
\expandafter\ifx\csname bibnamefont\endcsname\relax
  \def\bibnamefont#1{#1}\fi
\expandafter\ifx\csname bibfnamefont\endcsname\relax
  \def\bibfnamefont#1{#1}\fi
\expandafter\ifx\csname citenamefont\endcsname\relax
  \def\citenamefont#1{#1}\fi
\expandafter\ifx\csname url\endcsname\relax
  \def\url#1{\texttt{#1}}\fi
\expandafter\ifx\csname urlprefix\endcsname\relax\def\urlprefix{URL }\fi
\providecommand{\bibinfo}[2]{#2}
\providecommand{\eprint}[2][]{\url{#2}}

\bibitem[{\citenamefont{Roslund et~al.}(2013)\citenamefont{Roslund, Medeiros~de
  Ara\'ujo, Jiang, Fabre, and Treps}}]{Roslund2013}
\bibinfo{author}{\bibfnamefont{J.}~\bibnamefont{Roslund}},
  \bibinfo{author}{\bibfnamefont{R.}~\bibnamefont{Medeiros~de Ara\'ujo}},
  \bibinfo{author}{\bibfnamefont{S.}~\bibnamefont{Jiang}},
  \bibinfo{author}{\bibfnamefont{C.}~\bibnamefont{Fabre}}, \bibnamefont{and}
  \bibinfo{author}{\bibfnamefont{N.}~\bibnamefont{Treps}},
  \bibinfo{journal}{Nature Photonics, doi:10.1038/nphoton.2013.340}
  (\bibinfo{year}{2013}).

\bibitem[{\citenamefont{O'Brien and Akira~Furusawa}(2009)}]{obrien2009photonic}
\bibinfo{author}{\bibfnamefont{J.~L.} \bibnamefont{O'Brien}} \bibnamefont{and}
  \bibinfo{author}{\bibfnamefont{J.~V.} \bibnamefont{Akira~Furusawa}},
  \bibinfo{journal}{Nature Photonics} \textbf{\bibinfo{volume}{3}},
  \bibinfo{pages}{687} (\bibinfo{year}{2009}).

\bibitem[{\citenamefont{Walther et~al.}(2005)\citenamefont{Walther, Resch,
  Rudolph, Schenck, Weinfurter, Vedral, Aspelmeyer, and
  Zeilinger}}]{walther2005experimental}
\bibinfo{author}{\bibfnamefont{P.}~\bibnamefont{Walther}},
  \bibinfo{author}{\bibfnamefont{K.~J.} \bibnamefont{Resch}},
  \bibinfo{author}{\bibfnamefont{T.}~\bibnamefont{Rudolph}},
  \bibinfo{author}{\bibfnamefont{E.}~\bibnamefont{Schenck}},
  \bibinfo{author}{\bibfnamefont{H.}~\bibnamefont{Weinfurter}},
  \bibinfo{author}{\bibfnamefont{V.}~\bibnamefont{Vedral}},
  \bibinfo{author}{\bibfnamefont{M.}~\bibnamefont{Aspelmeyer}},
  \bibnamefont{and}
  \bibinfo{author}{\bibfnamefont{A.}~\bibnamefont{Zeilinger}},
  \bibinfo{journal}{Nature} \textbf{\bibinfo{volume}{434}},
  \bibinfo{pages}{169} (\bibinfo{year}{2005}).

\bibitem[{\citenamefont{Ukai et~al.}(2011)\citenamefont{Ukai, Iwata, Shimokawa,
  Armstrong, Politi, Yoshikawa, van Loock, and Furusawa}}]{Furusawa2011}
\bibinfo{author}{\bibfnamefont{R.}~\bibnamefont{Ukai}},
  \bibinfo{author}{\bibfnamefont{N.}~\bibnamefont{Iwata}},
  \bibinfo{author}{\bibfnamefont{Y.}~\bibnamefont{Shimokawa}},
  \bibinfo{author}{\bibfnamefont{S.~C.} \bibnamefont{Armstrong}},
  \bibinfo{author}{\bibfnamefont{A.}~\bibnamefont{Politi}},
  \bibinfo{author}{\bibfnamefont{J.-i.} \bibnamefont{Yoshikawa}},
  \bibinfo{author}{\bibfnamefont{P.}~\bibnamefont{van Loock}},
  \bibnamefont{and} \bibinfo{author}{\bibfnamefont{A.}~\bibnamefont{Furusawa}},
  \bibinfo{journal}{Phys. Rev. Lett.} \textbf{\bibinfo{volume}{106}},
  \bibinfo{pages}{240504} (\bibinfo{year}{2011}).

\bibitem[{\citenamefont{Kok et~al.}(2007)\citenamefont{Kok, Munro, Nemoto,
  Ralph, Dowling, and Milburn}}]{kok2007linear}
\bibinfo{author}{\bibfnamefont{P.}~\bibnamefont{Kok}},
  \bibinfo{author}{\bibfnamefont{W.~J.} \bibnamefont{Munro}},
  \bibinfo{author}{\bibfnamefont{K.}~\bibnamefont{Nemoto}},
  \bibinfo{author}{\bibfnamefont{T.~C.} \bibnamefont{Ralph}},
  \bibinfo{author}{\bibfnamefont{J.~P.} \bibnamefont{Dowling}},
  \bibnamefont{and} \bibinfo{author}{\bibfnamefont{G.}~\bibnamefont{Milburn}},
  \bibinfo{journal}{Rev. Mod. Phys.} \textbf{\bibinfo{volume}{79}},
  \bibinfo{pages}{135} (\bibinfo{year}{2007}).

\bibitem[{\citenamefont{Ralph and Pryde}(2010)}]{ralph2010optical}
\bibinfo{author}{\bibfnamefont{T.~C.} \bibnamefont{Ralph}} \bibnamefont{and}
  \bibinfo{author}{\bibfnamefont{G.~J.} \bibnamefont{Pryde}},
  \bibinfo{journal}{Progress in Optics} \textbf{\bibinfo{volume}{54}},
  \bibinfo{pages}{209} (\bibinfo{year}{2010}).

\bibitem[{\citenamefont{Knill et~al.}(2001)\citenamefont{Knill, Laflamme, and
  Milburn}}]{knill2001scheme}
\bibinfo{author}{\bibfnamefont{E.}~\bibnamefont{Knill}},
  \bibinfo{author}{\bibfnamefont{R.}~\bibnamefont{Laflamme}}, \bibnamefont{and}
  \bibinfo{author}{\bibfnamefont{G.~J.} \bibnamefont{Milburn}},
  \bibinfo{journal}{Nature} \textbf{\bibinfo{volume}{409}}, \bibinfo{pages}{46}
  (\bibinfo{year}{2001}).

\bibitem[{\citenamefont{Raussendorf and Briegel}(2001)}]{raussendorf2001one}
\bibinfo{author}{\bibfnamefont{R.}~\bibnamefont{Raussendorf}} \bibnamefont{and}
  \bibinfo{author}{\bibfnamefont{H.~J.} \bibnamefont{Briegel}},
  \bibinfo{journal}{Phys. Rev. Lett.} \textbf{\bibinfo{volume}{86}},
  \bibinfo{pages}{5188} (\bibinfo{year}{2001}).

\bibitem[{\citenamefont{Menicucci et~al.}(2006)\citenamefont{Menicucci, van
  Loock, Gu, Weedbrook, Ralph, and Nielsen}}]{Nielsen2006}
\bibinfo{author}{\bibfnamefont{N.~C.} \bibnamefont{Menicucci}},
  \bibinfo{author}{\bibfnamefont{P.}~\bibnamefont{van Loock}},
  \bibinfo{author}{\bibfnamefont{M.}~\bibnamefont{Gu}},
  \bibinfo{author}{\bibfnamefont{C.}~\bibnamefont{Weedbrook}},
  \bibinfo{author}{\bibfnamefont{T.~C.} \bibnamefont{Ralph}}, \bibnamefont{and}
  \bibinfo{author}{\bibfnamefont{M.~A.} \bibnamefont{Nielsen}},
  \bibinfo{journal}{Phys. Rev. Lett.} \textbf{\bibinfo{volume}{97}},
  \bibinfo{pages}{110501} (\bibinfo{year}{2006}).

\bibitem[{\citenamefont{Weedbrook et~al.}(2012)\citenamefont{Weedbrook,
  Pirandola, Garc\'{i}a-Patr\'{o}n, Cerf, Ralph, Shapiro, and
  Lloyd}}]{Lloyd2012}
\bibinfo{author}{\bibfnamefont{C.}~\bibnamefont{Weedbrook}},
  \bibinfo{author}{\bibfnamefont{S.}~\bibnamefont{Pirandola}},
  \bibinfo{author}{\bibfnamefont{R.}~\bibnamefont{Garc\'{i}a-Patr\'{o}n}},
  \bibinfo{author}{\bibfnamefont{N.~J.} \bibnamefont{Cerf}},
  \bibinfo{author}{\bibfnamefont{T.~C.} \bibnamefont{Ralph}},
  \bibinfo{author}{\bibfnamefont{J.~H.} \bibnamefont{Shapiro}},
  \bibnamefont{and} \bibinfo{author}{\bibfnamefont{S.}~\bibnamefont{Lloyd}},
  \bibinfo{journal}{Rev. Mod. Phys.} \textbf{\bibinfo{volume}{84}},
  \bibinfo{pages}{621} (\bibinfo{year}{2012}).

\bibitem[{\citenamefont{Yukawa et~al.}(2008)\citenamefont{Yukawa, Ukai, van
  Loock, and Furusawa}}]{vanLoock2008}
\bibinfo{author}{\bibfnamefont{M.}~\bibnamefont{Yukawa}},
  \bibinfo{author}{\bibfnamefont{R.}~\bibnamefont{Ukai}},
  \bibinfo{author}{\bibfnamefont{P.}~\bibnamefont{van Loock}},
  \bibnamefont{and} \bibinfo{author}{\bibfnamefont{A.}~\bibnamefont{Furusawa}},
  \bibinfo{journal}{Phys. Rev. A} \textbf{\bibinfo{volume}{78}},
  \bibinfo{pages}{012301} (\bibinfo{year}{2008}).

\bibitem[{\citenamefont{van Loock et~al.}(2007)\citenamefont{van Loock,
  Weedbrook, and Gu}}]{vanLoock2007}
\bibinfo{author}{\bibfnamefont{P.}~\bibnamefont{van Loock}},
  \bibinfo{author}{\bibfnamefont{C.}~\bibnamefont{Weedbrook}},
  \bibnamefont{and} \bibinfo{author}{\bibfnamefont{M.}~\bibnamefont{Gu}},
  \bibinfo{journal}{Phys. Rev. A} \textbf{\bibinfo{volume}{76}},
  \bibinfo{pages}{032321} (\bibinfo{year}{2007}).

\bibitem[{\citenamefont{Yokoyama et~al.}(2013)\citenamefont{Yokoyama, Ukai,
  Armstrong, Sornphiphatphong, Kaji, Suzuki, Yoshikawa, Yonezawa, Menicucci,
  and Furusawa}}]{yokoyama2013optical}
\bibinfo{author}{\bibfnamefont{S.}~\bibnamefont{Yokoyama}},
  \bibinfo{author}{\bibfnamefont{R.}~\bibnamefont{Ukai}},
  \bibinfo{author}{\bibfnamefont{S.~C.} \bibnamefont{Armstrong}},
  \bibinfo{author}{\bibfnamefont{C.}~\bibnamefont{Sornphiphatphong}},
  \bibinfo{author}{\bibfnamefont{T.}~\bibnamefont{Kaji}},
  \bibinfo{author}{\bibfnamefont{S.}~\bibnamefont{Suzuki}},
  \bibinfo{author}{\bibfnamefont{J.-i.} \bibnamefont{Yoshikawa}},
  \bibinfo{author}{\bibfnamefont{H.}~\bibnamefont{Yonezawa}},
  \bibinfo{author}{\bibfnamefont{N.~C.} \bibnamefont{Menicucci}},
  \bibnamefont{and} \bibinfo{author}{\bibfnamefont{A.}~\bibnamefont{Furusawa}},
  \bibinfo{journal}{Nature Photonics} \textbf{\bibinfo{volume}{7}},
  \bibinfo{pages}{982} (\bibinfo{year}{2013}).

\bibitem[{\citenamefont{Armstrong et~al.}(2012)\citenamefont{Armstrong,
  Morizur, Janousek, Hage, Treps, Lam, and Bachor}}]{Armstrong2012}
\bibinfo{author}{\bibfnamefont{S.}~\bibnamefont{Armstrong}},
  \bibinfo{author}{\bibfnamefont{J.-F.} \bibnamefont{Morizur}},
  \bibinfo{author}{\bibfnamefont{J.}~\bibnamefont{Janousek}},
  \bibinfo{author}{\bibfnamefont{B.}~\bibnamefont{Hage}},
  \bibinfo{author}{\bibfnamefont{N.}~\bibnamefont{Treps}},
  \bibinfo{author}{\bibfnamefont{P.~K.} \bibnamefont{Lam}}, \bibnamefont{and}
  \bibinfo{author}{\bibfnamefont{H.-A.} \bibnamefont{Bachor}},
  \bibinfo{journal}{Nature Commun.} \textbf{\bibinfo{volume}{3}},
  \bibinfo{pages}{1026} (\bibinfo{year}{2012}).

\bibitem[{\citenamefont{Pysher et~al.}(2011)\citenamefont{Pysher, Miwa,
  Shahrokhshahi, Bloomer, and Pfister}}]{pfister2011}
\bibinfo{author}{\bibfnamefont{M.}~\bibnamefont{Pysher}},
  \bibinfo{author}{\bibfnamefont{Y.}~\bibnamefont{Miwa}},
  \bibinfo{author}{\bibfnamefont{R.}~\bibnamefont{Shahrokhshahi}},
  \bibinfo{author}{\bibfnamefont{R.}~\bibnamefont{Bloomer}}, \bibnamefont{and}
  \bibinfo{author}{\bibfnamefont{O.}~\bibnamefont{Pfister}},
  \bibinfo{journal}{Phys. Rev. Lett.} \textbf{\bibinfo{volume}{107}},
  \bibinfo{pages}{030505} (\bibinfo{year}{2011}).

\bibitem[{\citenamefont{Menicucci et~al.}(2008)\citenamefont{Menicucci,
  Flammia, and Pfister}}]{pfister2008}
\bibinfo{author}{\bibfnamefont{N.~C.} \bibnamefont{Menicucci}},
  \bibinfo{author}{\bibfnamefont{S.~T.} \bibnamefont{Flammia}},
  \bibnamefont{and} \bibinfo{author}{\bibfnamefont{O.}~\bibnamefont{Pfister}},
  \bibinfo{journal}{Phys. Rev. Lett.} \textbf{\bibinfo{volume}{101}},
  \bibinfo{pages}{130501} (\bibinfo{year}{2008}).

\bibitem[{\citenamefont{Chen et~al.}(2013)\citenamefont{Chen, Menicucci, and
  Pfister}}]{Chen2013}
\bibinfo{author}{\bibfnamefont{M.}~\bibnamefont{Chen}},
  \bibinfo{author}{\bibfnamefont{N.~C.} \bibnamefont{Menicucci}},
  \bibnamefont{and} \bibinfo{author}{\bibfnamefont{O.}~\bibnamefont{Pfister}},
  \bibinfo{journal}{arXiv:1311.2957}  (\bibinfo{year}{2013}).

\bibitem[{\citenamefont{Wenger et~al.}(2004)\citenamefont{Wenger,
  Tualle-Brouri, and Grangier}}]{wenger2004non}
\bibinfo{author}{\bibfnamefont{J.}~\bibnamefont{Wenger}},
  \bibinfo{author}{\bibfnamefont{R.}~\bibnamefont{Tualle-Brouri}},
  \bibnamefont{and} \bibinfo{author}{\bibfnamefont{P.}~\bibnamefont{Grangier}},
  \bibinfo{journal}{Phys. Rev. Lett.} \textbf{\bibinfo{volume}{92}},
  \bibinfo{pages}{153601} (\bibinfo{year}{2004}).

\bibitem[{\citenamefont{Pinel et~al.}(2012)\citenamefont{Pinel, Jian,
  de~Ara\'ujo, Feng, Chalopin, Fabre, and Treps}}]{Pinel2012}
\bibinfo{author}{\bibfnamefont{O.}~\bibnamefont{Pinel}},
  \bibinfo{author}{\bibfnamefont{P.}~\bibnamefont{Jian}},
  \bibinfo{author}{\bibfnamefont{R.~M.} \bibnamefont{de~Ara\'ujo}},
  \bibinfo{author}{\bibfnamefont{J.}~\bibnamefont{Feng}},
  \bibinfo{author}{\bibfnamefont{B.}~\bibnamefont{Chalopin}},
  \bibinfo{author}{\bibfnamefont{C.}~\bibnamefont{Fabre}}, \bibnamefont{and}
  \bibinfo{author}{\bibfnamefont{N.}~\bibnamefont{Treps}},
  \bibinfo{journal}{Phys. Rev. Lett.} \textbf{\bibinfo{volume}{108}},
  \bibinfo{pages}{083601} (\bibinfo{year}{2012}).

\bibitem[{\citenamefont{Grice et~al.}(2001)\citenamefont{Grice, URen, and
  Walmsley}}]{walmsley2001}
\bibinfo{author}{\bibfnamefont{W.~P.} \bibnamefont{Grice}},
  \bibinfo{author}{\bibfnamefont{A.~B.} \bibnamefont{URen}}, \bibnamefont{and}
  \bibinfo{author}{\bibfnamefont{I.~A.} \bibnamefont{Walmsley}},
  \bibinfo{journal}{Phys. Rev. A} \textbf{\bibinfo{volume}{64}},
  \bibinfo{pages}{063815} (\bibinfo{year}{2001}).

\bibitem[{\citenamefont{Mosley et~al.}(2008)\citenamefont{Mosley, Lundeen,
  Smith, Wasylczyk, URen, Silberhorn, and Walmsley}}]{walmsley2008}
\bibinfo{author}{\bibfnamefont{P.~J.} \bibnamefont{Mosley}},
  \bibinfo{author}{\bibfnamefont{J.~S.} \bibnamefont{Lundeen}},
  \bibinfo{author}{\bibfnamefont{B.~J.} \bibnamefont{Smith}},
  \bibinfo{author}{\bibfnamefont{P.}~\bibnamefont{Wasylczyk}},
  \bibinfo{author}{\bibfnamefont{A.~B.} \bibnamefont{URen}},
  \bibinfo{author}{\bibfnamefont{C.}~\bibnamefont{Silberhorn}},
  \bibnamefont{and} \bibinfo{author}{\bibfnamefont{I.~A.}
  \bibnamefont{Walmsley}}, \bibinfo{journal}{Phys. Rev. Lett.}
  \textbf{\bibinfo{volume}{100}}, \bibinfo{pages}{133601}
  (\bibinfo{year}{2008}).

\bibitem[{\citenamefont{Patera et~al.}(2010)\citenamefont{Patera, Treps, Fabre,
  and de~Valcarcel}}]{patera2010}
\bibinfo{author}{\bibfnamefont{G.}~\bibnamefont{Patera}},
  \bibinfo{author}{\bibfnamefont{N.}~\bibnamefont{Treps}},
  \bibinfo{author}{\bibfnamefont{C.}~\bibnamefont{Fabre}}, \bibnamefont{and}
  \bibinfo{author}{\bibfnamefont{G.~J.} \bibnamefont{de~Valcarcel}},
  \bibinfo{journal}{Eur. Phys. J. D} \textbf{\bibinfo{volume}{56}},
  \bibinfo{pages}{123} (\bibinfo{year}{2010}).

\bibitem[{\citenamefont{de~Valcarcel et~al.}(2006)\citenamefont{de~Valcarcel,
  Patera, Treps, and Fabre}}]{Valcarcel12}
\bibinfo{author}{\bibfnamefont{G.~J.} \bibnamefont{de~Valcarcel}},
  \bibinfo{author}{\bibfnamefont{G.}~\bibnamefont{Patera}},
  \bibinfo{author}{\bibfnamefont{N.}~\bibnamefont{Treps}}, \bibnamefont{and}
  \bibinfo{author}{\bibfnamefont{C.}~\bibnamefont{Fabre}},
  \bibinfo{journal}{Phys. Rev. A} \textbf{\bibinfo{volume}{74}},
  \bibinfo{pages}{061801(R)} (\bibinfo{year}{2006}).

\bibitem[{\citenamefont{Opatrn\'{y} et~al.}(2002)\citenamefont{Opatrn\'{y},
  Korolkova, and Leuchs}}]{Leuchs2002}
\bibinfo{author}{\bibfnamefont{T.}~\bibnamefont{Opatrn\'{y}}},
  \bibinfo{author}{\bibfnamefont{N.}~\bibnamefont{Korolkova}},
  \bibnamefont{and} \bibinfo{author}{\bibfnamefont{G.}~\bibnamefont{Leuchs}},
  \bibinfo{journal}{Phys. Rev. A} \textbf{\bibinfo{volume}{66}},
  \bibinfo{pages}{053813} (\bibinfo{year}{2002}).

\bibitem[{\citenamefont{Braunstein and van Loock}(2005)}]{Braunstein2005}
\bibinfo{author}{\bibfnamefont{S.~L.} \bibnamefont{Braunstein}}
  \bibnamefont{and} \bibinfo{author}{\bibfnamefont{P.}~\bibnamefont{van
  Loock}}, \bibinfo{journal}{Rev. Mod. Phys.} \textbf{\bibinfo{volume}{77}},
  \bibinfo{pages}{513} (\bibinfo{year}{2005}).

\bibitem[{\citenamefont{Menicucci et~al.}(2011)\citenamefont{Menicucci,
  Flammia, and van Loock}}]{menicucci2011graphical}
\bibinfo{author}{\bibfnamefont{N.~C.} \bibnamefont{Menicucci}},
  \bibinfo{author}{\bibfnamefont{S.~T.} \bibnamefont{Flammia}},
  \bibnamefont{and} \bibinfo{author}{\bibfnamefont{P.}~\bibnamefont{van
  Loock}}, \bibinfo{journal}{Physical Review A} \textbf{\bibinfo{volume}{83}},
  \bibinfo{pages}{042335} (\bibinfo{year}{2011}).

\bibitem[{\citenamefont{Dutta et~al.}(1995)\citenamefont{Dutta, Mukunda, Simon
  et~al.}}]{dutta1995real}
\bibinfo{author}{\bibfnamefont{B.}~\bibnamefont{Dutta}},
  \bibinfo{author}{\bibfnamefont{N.}~\bibnamefont{Mukunda}},
  \bibinfo{author}{\bibfnamefont{R.}~\bibnamefont{Simon}},
  \bibnamefont{et~al.}, \bibinfo{journal}{Pramana}
  \textbf{\bibinfo{volume}{45}}, \bibinfo{pages}{471} (\bibinfo{year}{1995}).

\bibitem[{\citenamefont{Ferrini et~al.}(2013)\citenamefont{Ferrini, Gazeau,
  Coudreau, Fabre, and Treps}}]{ferrini2013compact}
\bibinfo{author}{\bibfnamefont{G.}~\bibnamefont{Ferrini}},
  \bibinfo{author}{\bibfnamefont{J.~P.} \bibnamefont{Gazeau}},
  \bibinfo{author}{\bibfnamefont{T.}~\bibnamefont{Coudreau}},
  \bibinfo{author}{\bibfnamefont{C.}~\bibnamefont{Fabre}}, \bibnamefont{and}
  \bibinfo{author}{\bibfnamefont{N.}~\bibnamefont{Treps}},
  \bibinfo{journal}{New J. Phys.} \textbf{\bibinfo{volume}{15}},
  \bibinfo{pages}{093015} (\bibinfo{year}{2013}).

\bibitem[{\citenamefont{Vaughan et~al.}(2005)\citenamefont{Vaughan, Hornung,
  Feurer, and Nelson}}]{nelson2005}
\bibinfo{author}{\bibfnamefont{J.}~\bibnamefont{Vaughan}},
  \bibinfo{author}{\bibfnamefont{T.}~\bibnamefont{Hornung}},
  \bibinfo{author}{\bibfnamefont{T.}~\bibnamefont{Feurer}}, \bibnamefont{and}
  \bibinfo{author}{\bibfnamefont{K.}~\bibnamefont{Nelson}},
  \bibinfo{journal}{Opt. Lett.} \textbf{\bibinfo{volume}{30}},
  \bibinfo{pages}{323} (\bibinfo{year}{2005}).

\bibitem[{\citenamefont{Simon}(2000)}]{simon2000}
\bibinfo{author}{\bibfnamefont{R.}~\bibnamefont{Simon}},
  \bibinfo{journal}{Phys. Rev. Lett.} \textbf{\bibinfo{volume}{84}},
  \bibinfo{pages}{2726} (\bibinfo{year}{2000}).

\bibitem[{\citenamefont{Braunstein}(2005)}]{Braunstein2005-irreducible}
\bibinfo{author}{\bibfnamefont{S.~L.} \bibnamefont{Braunstein}},
  \bibinfo{journal}{Phys. Rev. A} \textbf{\bibinfo{volume}{71}},
  \bibinfo{pages}{055801} (\bibinfo{year}{2005}).

\bibitem[{\citenamefont{Roslund et~al.}(2009)\citenamefont{Roslund, Shir,
  B{\"a}ck, and Rabitz}}]{roslund2009accelerated}
\bibinfo{author}{\bibfnamefont{J.}~\bibnamefont{Roslund}},
  \bibinfo{author}{\bibfnamefont{O.~M.} \bibnamefont{Shir}},
  \bibinfo{author}{\bibfnamefont{T.}~\bibnamefont{B{\"a}ck}}, \bibnamefont{and}
  \bibinfo{author}{\bibfnamefont{H.}~\bibnamefont{Rabitz}},
  \bibinfo{journal}{Physical Review A} \textbf{\bibinfo{volume}{80}},
  \bibinfo{pages}{043415} (\bibinfo{year}{2009}).

\end{thebibliography}

\end{document}